\documentclass[final,5p,times,twocolumn]{elsarticle}
\usepackage[utf8]{inputenc}
\usepackage[T1]{fontenc}
\usepackage{orcidlink}
\usepackage{hyperref}
\usepackage{amssymb}
\usepackage{lipsum}
\usepackage{siunitx}
\usepackage{textgreek}
\usepackage[super]{nth}
\usepackage[version=4]{mhchem}
\usepackage{lineno}

\DeclareSIUnit\bar{bar}

\journal{Physics Letters B}

\begin{document}
\begin{frontmatter}

\title{Further search for magnetic-field-induced neutron disappearance in an ultracold neutron beam}

\affiliation[inst2]{organization={Universit\'e Grenoble Alpes, CNRS, Grenoble INP, LPSC-IN2P3}, city={Grenoble}, postcode={38026}, country={France}}
\affiliation[inst4]{organization={Institut Laue-Langevin}, addressline={71 Avenue des Martyrs}, city={Grenoble}, postcode={38042}, country={France}}
\affiliation[inst1]{organization={Normandie Universit\'e, ENSICAEN, UNICAEN, CNRS/IN2P3, LPC Caen}, city={Caen}, postcode={14000}, country={France}}
\affiliation[inst3]{organization={Sorbonne Universit\'e, CNRS/IN2P3, Laboratoire de Physique Nucl\'eaire et de Hautes Energies (LPNHE)}, city={Paris}, postcode={75005}, country={France}}

\author[inst2]{Gaby Brenot}
\author[inst2]{Benoit Cl\'ement\orcidlink{https://orcid.org/0000-0002-0079-6104}\corref{cor2}}
\author[inst4]{Hanno Filter-Pieler\orcidlink{https://orcid.org/0000-0001-5627-3807}}
\author[inst1]{Daniel Galbinski\orcidlink{https://orcid.org/0009-0002-0214-2064}\corref{cor1}}
\author[inst4]{Tobias Jenke\orcidlink{https://orcid.org/0000-0002-7815-4726}}
\author[inst1]{Thomas Lefort\orcidlink{https://orcid.org/0000-0003-2198-2093}}
\author[inst1]{Anthony Lejuez}
\author[inst2]{Guillaume Pignol\orcidlink{https://orcid.org/0000-0001-7086-0100}}
\author[inst2]{Stephanie Roccia\orcidlink{https://orcid.org/0009-0004-4752-5442}}
\author[inst3]{William Saenz-Arevalo\orcidlink{https://orcid.org/0009-0005-1399-9524}}

\cortext[cor1]{galbinski@lpccaen.in2p3.fr}
\cortext[cor2]{bclement@lpsc.in2p3.fr}

\begin{abstract}
We report the results of the second iteration of an experiment searching for neutron-hidden-neutron oscillations in a beam of ultracold neutrons, 
conducted at the PF2 facility of the Institut Laue Langevin (ILL). 
Oscillations were tested via neutron disappearance as a function of an applied magnetic field, 
in the context of a phenomenological two-parameter model assuming zero hidden potentials. 
The magnetic field was varied in a step-wise manner in order to resonantly enhance the oscillation probability 
at different mass splittings ($\delta m$) across a $60$--$1550$\,\si{\pico\electronvolt} range. 
No evidence for neutron disappearance is observed and conservative limits on the neutron-hidden-neutron oscillation period ($\tau_{nn'}$) have been set 
at 95\,\% confidence level: \mbox{$\tau_{nn'} > 200$\,\si{\milli\second}} for \mbox{$|\delta m| \in [60, 400]$\,\si{\pico\electronvolt}} 
and \mbox{$\tau_{nn'} > 100$\,\si{\milli\second}} for \mbox{$|\delta m| \in [400, 1550]$\,\si{\pico\electronvolt}}.
\end{abstract}


\end{frontmatter}

\section{Introduction}

A large number of Standard Model (SM) extensions include what are commonly referred to as ``hidden sectors'', typically involving new particles (and gauge interactions) that couple to SM particles only via gravity and so-called ``portals'' which facilitate mixing between the two sectors. Many of these models postulate a ``mirror'' duplicate of the Standard Model (MSM) with the same $\textit{SU}(3)$ $\times$ $\textit{SU}(2)$ $\times$ $\textit{U}(1)$ gauge symmetry, in which every particle has the opposite chirality to its SM counterpart \cite{Berezhiani2004,Sarrazin2012,Okun2007}. This idea was first proposed in 1956 \cite{Lee1956} in the context of the $\tau$-$\theta$ puzzle and subsequently expanded upon \cite{Okun1966, Blinnikov1983, Foot1991} in the decades after the discovery of parity violation \cite{Wu1957}. Alternatively, there exist superstring-based models in which ``ordinary'' and ``hidden'' particles exist in parallel branes of a higher-dimensional space (bulk) \cite{Das2011,Dvali2009,Arkani-Hamed2000,Brax2004}, where they exert gravitational influence on each other that is interpreted in our brane as dark matter. 

Hidden-sector models formulated with mirror matter have gained particular attention for their ability to address a number of outstanding problems in fundamental physics. Whilst the usual weak interaction has a vector-axial-vector ($V-A$) structure, such that parity is maximally violated, mirror particles experience analogous $V+A$ interactions and are sterile under all SM interactions. The combined Lagrangian $L = L_{\text{SM}} + L_{\text{MSM}}$ is then invariant under parity transformations, thus restoring global parity symmetry. Furthermore, it has been proposed that the mirror sector could provide suitable candidates for the hypothesised non-baryonic (dark matter) component of the universe's matter content  \cite{Foot2004,Foot2004v2,Ignatiev2003,Blinnikov2010}. The presence of mirror particles would have also impacted the evolution of the early universe \cite{Berezhiani2001,Foot2014,Berezhiani2005,Coc2013,Coc2014}, including large-scale structure formation and Big Bang nucleosynthesis, as well as certain astrophysical phenomena \cite{Berezhiani2012}.

Experimentally accessible particle-hidden-particle interactions include oscillations of electrically neutral particles such as photons, neutrons, and neutrinos, assuming some type of non-gravitational coupling. Incidentally, such transitions can readily yield a mechanism for baryon number violation \cite{Bento2001,Berezhiani2018v2}, one of the requirements to explain the matter-antimatter asymmetry of the universe \cite{Canetti2012,Sakharov1991}. This work focuses on neutron-hidden-neutron \mbox{($n-n'$)} oscillations, which have been studied extensively in the past two decades \cite{Tan2019,Berezhiani2006,Berezhiani2019,Mohapatra2005} from both experimental and theoretical viewpoints. We consider a generic two-state model in which the Hamiltonian describing \mbox{$n-n'$} mixing may be expressed as (see for example Refs.\@ \cite{Berezhiani2009,Hostert2023}):

\begin{equation}
    H = \begin{pmatrix}
         m_{n} + \Delta E & \epsilon_{nn'} \\
         \epsilon_{nn'} & m_n + \delta m + \Delta E' 
         \end{pmatrix} \:,
\end{equation}

where $m_n$ is the neutron mass, $\Delta E$ ($\Delta E'$) is the energy of the neutron (hidden neutron) from interactions with its environment, $\delta m$ is the neutron-hidden-neutron mass splitting which could arise from spontaneously broken $\mathbb{Z}_2$ symmetry between the ordinary and hidden sectors \cite{Mohapatra2018,Kamyshkov2022}, and $\epsilon_{nn'}$ is a mixing parameter whose inverse $\tau_{nn'} = \epsilon^{-1}_{nn'}$ determines the timescale of the oscillations.

By assuming zero hidden potentials ($\Delta E' = 0)$, it follows from the solution of the Schr\"{o}dinger equation that the neutron-hidden-neutron transition probability, as a function of free-flight time $t_f$, is given by:

\begin{equation}\label{eq:prob}
    P_{nn'}(t_f) = \sin^2{(2\theta)} \sin{\Bigg(\frac{1}{2}t_f\sqrt{(\Delta E - \delta m)^2 + 4 \epsilon^2_{nn'}}\Bigg)} \:,
\end{equation}

where the mixing amplitude $\theta$ is defined as:

\begin{equation}\label{eq:angle}
    \sin^2{(2\theta)} = \frac{4 \epsilon^2_{nn'}}{4 \epsilon^2_{nn'} + (\Delta E - \delta m)^2} \:.
\end{equation}

Equation \ref{eq:prob} is valid from the time of a free neutron's last scatter (or creation) at $t_f = 0$, until its next scatter (or decay). For a given $\delta m$ and $\epsilon_{nn'}$, the transition probability is maximised when the resonance condition $\Delta E = \delta m$ is fulfilled.

Neutron-hidden-neutron oscillations have been searched for in laboratory experiments through two main channels: anomalous disappearance of trapped ultracold neutrons (UCNs) as a function of an applied magnetic field \cite{Ban2007,Serebrov2009,Altarev2009,Berezhiani2012v2,Berezhiani2018,Ayres2022} and passing-through-wall neutron regeneration experiments \cite{Stasser2021,Almazn2022,Broussard2022,Gonzalez2024} where the relevant signature is an excess neutron flux on the other side of a potential barrier (e.g.\@ dense neutron shielding). This article presents the results of a second experiment probing \mbox{$n-n'$} oscillations via neutron disappearance in an unpolarised UCN beam, following Ref.\@ \cite{saenz}, exploring the poorly constrained mass splitting regime of $60$--$1550$\,\si{\pico\electronvolt}. As previously, the transition probability was enhanced by application of an external magnetic field to satisfy the resonance condition at different $\delta m$, giving the neutrons a maximum energy shift equal to $\Delta E = \mu_n B$ (where $\mu_n = -60.3$\,\si{\pico\electronvolt\per\milli\tesla} is the neutron magnetic moment).

\section{Experimental Setup}

\begin{figure*}
    \centering
    \includegraphics[width=1.0\linewidth]{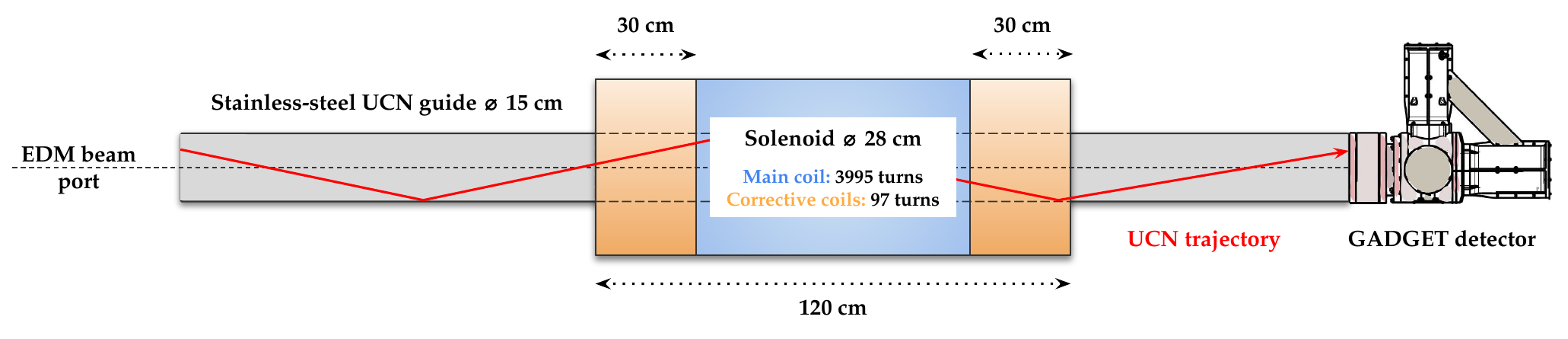}
    \vspace{-5mm}
    \caption{Schematic of the experimental setup at the PF2 ultracold neutron facility. The red arrow represents a possible UCN trajectory from the beam port to the detector. Note that the stated dimensions are not to scale.}
    \label{fig:setup}
    \vspace{-4mm}
\end{figure*}

The experiment\@ \cite{pf2} was conducted in May--July 2024 at the PF2 ultracold neutron facility of the Institut Laue Langevin (ILL) in Grenoble, France. The salient components of the PF2 facility are the liquid \ce{D_2} cold source, curved vertical extraction guide, and Steyerl turbine\@ \cite{Steyerl1975}, which together create a total UCN flux of around \SI{2.6E4}{\per\centi\metre\squared\per\second} (up to date as of Ref.\@ \cite{steyerl1987}) for vertical velocities less than 6.2\,\si{\metre\per\second} (\SI{201}{\nano\electronvolt}). A schematic of the experiment apparatus is displayed in Fig.\@ \ref{fig:setup}. UCNs from the ``EDM'' beam port are transported along an evacuated horizontal 15\,\si{\centi\metre} diameter uncoated stainless-steel guide towards a \SI{1.2}{\metre} solenoid, which was re-purposed from another experiment \cite{coilthesis}. The solenoid was designed to generate a homogenous magnetic field up to \SI{30}{\milli\tesla} and has a main coil of 3995 turns (in 10 layers) of \SI{3.1}{\milli\metre} enamelled aluminium wire wrapped around an aluminium cylinder, with an inner bore diameter of \SI{28}{\centi\metre}. Two \SI{30}{\centi\metre} end-correction coils add an additional 97 turns each to improve the field uniformity. For this experiment, an independent on-site field mapping was performed to characterise the magnetic field inside the solenoid. 

Downstream of the solenoid is the UCN detector known as \mbox{GADGET}\@ \cite{saenz2}, as built for the n2EDM experiment\@ \cite{n2edm}. \mbox{GADGET} is a fast-response neutron counter based on a combination of two gases: \ce{^3He} for neutron capture and \ce{CF_4} for scintillation. The latter was selected because of its high photon yield, transparency to its own scintillation light (emitted in the 200--800\,\si{\nano\metre} range), and relatively low cost. Moreover, the rapid \ce{CF_4} decay time of approximately \SI{6}{\nano\second} \cite{Lehaut2015}, allows \mbox{GADGET} to handle counting rates up to a few $10^6$\,\si{\per\second} without significant pile-up. The gas mixture is contained in a cylindrical stainless steel chamber with radius \SI{7.4}{\centi\metre} and depth \SI{8.9}{\centi\metre}. The unpolarised UCN beam enters the detector through a circular window covered by a \SI{60}{\micro\metre} thick sheet of \ce{Al_{97}Mg_3} foil, which is mechanically supported by a stainless-steel grid.

Neutrons are captured on \ce{^3He} and the resulting disintegration products, a \SI{570}{\kilo\electronvolt} proton and a \SI{190}{\kilo\electronvolt} triton, provoke scintillation of the \ce{CF_4}. The chamber was filled with \ce{^3He} at \SI{25}{\milli\bar} followed by \SI{1015}{\milli\bar} of \ce{CF_4}. At these pressures, the UCN beam is entirely stopped in the gas mixture with a mean free path of \mbox{$\mathcal{O}$(1)\,\si{\centi\metre}}. In addition, SRIM simulations \cite{Ziegler2010} indicate that the corresponding proton (triton) range is less than \SI{5}{\milli\metre} (\SI{2}{\milli\metre}), meaning that in the majority of cases the total kinetic energy is fully converted to scintillation light. The light is collected by three Hamamatsu H3177-51 photomultipliers (PMTs) arranged along three mutually perpendicular axes around the detector chamber (one of which is parallel to the UCN beam). The PMTs are mounted at \SI{6}{\centi\metre} diameter quartz windows set in the chamber walls and held in place by supporting rods. A small amount of optical grease ensures optical contact between the PMT glass and the window. The inner walls of the chamber are polished to improve the reflection of incident photons towards the photosensors. 

Analogue signals from the PMTs were processed by FASTER (Fast Acquisition SysTem for nucleEar Research \cite{FASTER}), a highly modular DAQ system developed at LPC Caen. A \mbox{SYROCO\_AMC\_C5} motherboard synchronises the input channels via a master clock and houses three field-programmable gate arrays (FPGAs). The FPGAs perform online processing and triggering of the PMT waveforms, which are digitised by 12-bit analogue-to-digital converters with a \SI{500}{\mega\hertz} sampling rate. If all three PMTs are triggered within a rolling \SI{50}{\nano\second} window, the corresponding waveforms are recorded. This triple coincidence is imposed to remove dark counts from thermionic currents and only select events originating within the chamber itself (e.g.\@ UCN capture). The coincidence rate from uncorrelated signals is negligible. Instead of the full waveform information, only key parameters of individual waveforms, such as charge and amplitude, were written to disk. Further data reduction was achieved by saving exactly one in every ten events.


\section{Data-Taking Strategy}\label{sec:data-taking}

Two servers were run in parallel during data-taking. The FASTER DAQ was controlled by one server and the other hosted three threads for a backup acquisition\@ \cite{bourrion16} (counts only), a spectrum ADC card, and the solenoid power supplies. A large fraction of the measurements took place concurrently with another experiment at an adjacent beam port, so the two beams operated in ``time-shared'' mode where UCNs were alternately available every \SI{400}{\second} for \SI{200}{\second} delivery cycles. The cycles were divided into three successive analysing periods with coil-current values $I_A$, $I_B$, and $I_C$, where $I_A = I_B - 30$\,\si{\milli\ampere} and $I_C = I_B + 30$\,\si{\milli\ampere}, corresponding to variations of $\pm 12$\,\si{\micro\tesla}. 

The start of a cycle was marked by a TTL signal sent from the PF2 controller, signifying that the turbine is in position and the port shutter is open. Approximately 10 seconds later, a second TTL signal was generated by the spectrum ADC card in order to be able to synchronise the FASTER data with the solenoid field during offline analysis. Starting from the timestamp of the second TTL signal, the solenoid field was successively ramped from one field value to the next, according to the sequence described in Tab.\@ \ref{tab:sequence}.

\begin{table}[h!]
\normalsize
\renewcommand{\arraystretch}{1.2}
    \centering
    \begin{tabular}{cl}\hline\hline
        Time (\si{\second}) & Occurrence \\\hline
        $0$ & TTL synchronisation signal \\
        $18$--$21$ & Ramp solenoid to $A$-period field \\
        $21$--$62$ & Count UCNs at $A$-period field for \SI{41}{\second} \\
        $62$--$64$ & Ramp solenoid to $B$-period field \\
        $64$--$146$ & Count UCNs at $B$-period field for \SI{82}{\second} \\
        $146$--$149$ & Ramp solenoid to $C$-period field \\
        $149$--$190$ & Count UCNs at $C$-period field for \SI{41}{\second} \\\hline\hline
    \end{tabular}
    \caption{Time sequence of UCN counting and solenoid field ramping for standard data-taking cycles, starting from the second TTL signal.}
    \label{tab:sequence}
\end{table}

Overall, the UCN beam was allowed to stabilise for 21 seconds plus the time between the two TTL signals, in order to account for the gradual increase in event rate from the start of the cycle caused by the low energy component of the UCN spectrum. The time intervals allocated for field changes were of sufficient duration for the power supply to reach its target value and for UCNs that experienced the previous field to exit the solenoid volume. For each cycle, a self-normalising ratio sensitive to \mbox{$n-n'$} oscillations was constructed, defined as:

\begin{equation}
    R_{ABC} = \frac{N_B}{N_A + N_C} \:,
\end{equation}

where $N_A$, $N_B$, and $N_C$ are the integrated UCN counts during the $A$, $B$, and $C$ periods, respectively. The ratio is insensitive to long-term linear variations of the UCN rate due to changes in reactor power, turbine speed, and cold-source temperature, which occur over timescales much larger than a single UCN cycle. As such, the expected value of $R_{ABC}$ in the absence of \mbox{$n-n'$} oscillations is exactly unity, whereas it is less than one if the $B$-period field is $\approx \delta m / \mu_n$ and greater than one if the same is true of the $A$- or $C$-period fields.

To test for \mbox{$n-n'$} oscillations over a wide range of $\delta m$ values, a two-phase approach covering a low and a high energy regime was adopted. In the first phase (30 days), the coil field corresponding to period $B$ was scanned eight times between 0.28--2.12\,\si{\ampere} (60--500\,\si{\pico\electronvolt}) in steps of \SI{2.29}{\milli\ampere} (\SI{0.55}{\pico\electronvolt}). In the second phase (14 days), the coil field was scanned four times between 1.999--6.6\,\si{\ampere} (470--1550\,\si{\pico\electronvolt}) in steps of \SI{6.60}{\milli\ampere} (\SI{1.55}{\pico\electronvolt}). The direction of the solenoid current was periodically reversed such that half of the scans correspond to positive polarity and the other half to negative polarity.

Instead of a uniform field, a magnetic gradient, uniform over approximately 0.3\,\si{\metre}, was created by running a current through one of the correction coils, equal (for $I>2.12$\,\si{\ampere}) or slightly superior (otherwise) to the main coil current. This widened the oscillation resonance width at the cost of a reduced sensitivity, allowing us to a) cover a larger $\delta m$ range within the experimental time constraints and b) ensure that no $\delta m$ values were missed due to the finite resolution of the power supply (which also limited the step size). For \mbox{$I = 0.28$\,\si{\ampere}}, the magnetic field and gradient in the centre of the coil were recorded as \mbox{$B=1.1$\,\si{\milli\tesla}} and \mbox{$G=0.06$\,\si{\milli\tesla\per\metre}}, respectively, increasing to \SI{8.5}{\milli\tesla} (\SI{26.4}{\milli\tesla}) and \SI{0.13}{\milli\tesla\per\metre} (\SI{0.31}{\milli\tesla\per\metre}) at \mbox{$I=2.12$\,\si{\ampere}} (\mbox{$I=6.6$\,\si{\ampere}}).

In addition, a fluxgate situated below the solenoid was used to monitor the variations of the ambient magnetic field during the experiment, allowing field fluctuations (arising from power supply instabilities or movement of the ILL overhead crane) to be identified. It also recorded cycle-to-cycle magnetic variations with respect to a reference field measured at the start of data-taking with minimal current in the coils; two such references were created, one for each polarity of the main field.

The final dataset corresponds to roughly \SI{505}{\hour} beam-on exposure time across \num{9323} UCN delivery cycles and \SI{36}{\hour} beam-off data for background studies. \num{625} cycles ($6.7$\,\%) were discarded because of poor magnetic conditions. A further 13 cycles exhibited extreme outliers in event counts not consistent with the reactor power and the data files of 25 were incomplete or corrupted due to operational issues with the data acquisition; these 38 cycles were also rejected. The mean rate of beam-on FASTER coincidence events (after data quality checks) was around \num{2.8e5}\,\si{\per\second} at an average reactor power of \SI{49.9}{\mega\watt}. 

\section{Event Selection}

During beam-on data-taking, whilst most events correspond to UCN capture on \ce{^3He} inside the detector, there is a non-negligible background component that must be characterised and accounted for in offline analysis. We differentiate between two types of background: beam-induced and environmental. Environmental backgrounds are independent of the UCN beam and include fast neutrons, reactor \textgamma{}-rays, and cosmogenic muons. Interactions of \textgamma{}-rays in materials, such as the quartz windows or PMT glass, can liberate electrons that emit Cherenkov radiation as they travel through the medium. Furthermore, \textgamma{}-rays and muons can generate scintillation light by directly exciting or ionising the \ce{CF_4}. The total rate of the environmental backgrounds at the experiment site is estimated to be ${\mathcal{O}(20)}$ counts per second, which is statistically insignificant compared to the beam-on event rate.

\begin{table}
\footnotesize      
\renewcommand{\arraystretch}{1.3}
    \centering
    \begin{tabular}{lcc}\hline\hline
        \textbf{Variable} & \textbf{Description} & \textbf{Selection Cut} \\\hline
        (1) $q_i$ & Charge in PMT $i$ & $q_i \geq 400$\,a.u.\@ ($i = 1,2,3$) \\
        (2) $Q = \sum_{i=1}^{3} q_i$ & Summed Charge & $Q \geq 2367.75$\,a.u. \\
        (3) $A = \sum_{i=1}^{3} a_i$ & Summed Amplitude & $217.15\,\si{\milli\volt} \leq A \leq 750$\,\si{\milli\volt} \\
        (4) $Q / A$ & Charge over Amplitude & $7.21$\,a.u. $\leq Q / A \leq 13$\,a.u. \\
        (5) $D_1 = 1 - q_1 / Q$ & Relative Charge & $0.41717 \leq D_1 \leq 0.94206$ \\
        (6) $D_2 = 1 - q_2 / Q$ & Relative Charge & $0.37329 \leq D_2 \leq 0.94440$ \\
        (7) $D_3 = 1 - q_3 / Q$ & Relative Charge & $0.38473 \leq D_3 \leq 0.94056$ \\\hline\hline
    \end{tabular}
    \caption{Summary of the UCN selection cuts on pulse-shape parameters.}
    \label{tab:cuts}
\end{table}

Beam-induced backgrounds are primarily caused by neutron activation in the detector materials, releasing ionising \textbeta{} and \textgamma{} radiation that can traverse the gas mixture within the detector and thus satisfy the coincidence condition by illuminating all three PMTs near-simultaneously. The two most relevant \textbeta{}-decaying isotopes from activation in the detector materials are \ce{^{28}Al} in the entrance foil ($T_{1/2} = 2.245$\,\si{\minute} \cite{A28}) and \ce{^{56}Mn} in the chamber walls and supporting grid ($T_{1/2} = 154.7$\,\si{\minute} \cite{A56}), with event rates on the order of 10\,\% of the total UCN rate. Absorption by \ce{^{20}F} in the \ce{CF_4} and subsequent decay ($T_{1/2} = 11.07$\,\si{\second} \cite{A20}) is also possible, however, the rate of such events is inconsequential given that the probability of occurrence is modelled as over 180 times smaller than \ce{^3He} capture and around 20 times smaller than \ce{^{20}F} upscattering \cite{saenz3}.



A dedicated background sample was created following the completion of the final UCN measurement by taking data at the same location for around \SI{36}{\hour} without requesting the UCN beam. For the first 8 hours or so, the sample is dominated by events produced through \textbeta{}-decay and by environmental background from then on. Based on comparisons between statistically equivalent samples of beam-on and background events, as well as insights gained during and since the previous experiment \cite{saenz3}, a set of rectangular cuts on pulse-shape parameters has been devised to suppress the background rate to a minimal level. The cuts, outlined in Tab.\@ \ref{tab:cuts}, were initially optimised to maximise the $S_{12}$ significance metric \cite{Bityukov1998} via random grid search, but then modified manually to remove background more harshly and further reduce the rates of specific backgrounds not well handled by the optimisation.

\begin{figure}
    \centering
    \includegraphics[width=0.95\linewidth]{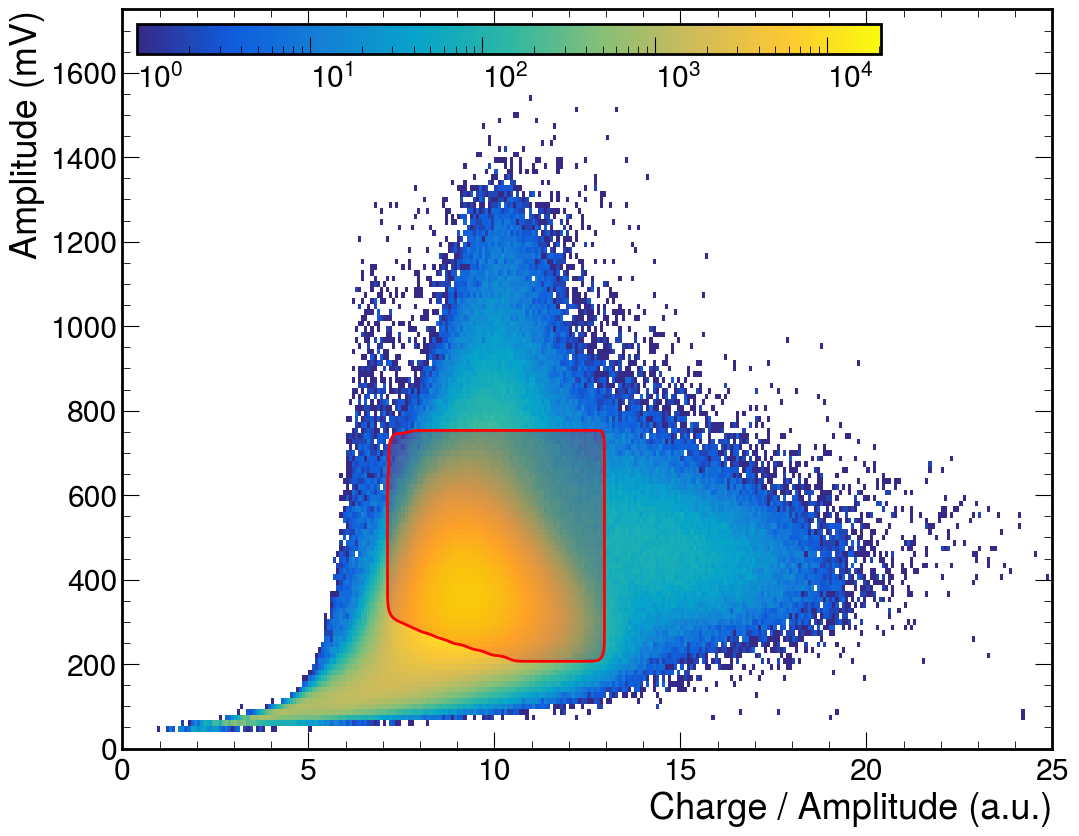}
        \vspace{-2mm}

    \caption{Pulse-shape analysis histogram ($A$ vs $Q / A$) binned from the data of representative UCN delivery cycles comprising \num{11357607} events. The red contour denotes the boundary of the smallest (Gaussian-smoothed) region containing all non-empty bins, after application of selection cuts 1--7. Approximately 32\,\% of rejected events (i.e.\@ from cuts 1 and 5--7) are found within this region.}
    \label{fig:psa}
    \vspace{-4mm}
\end{figure}

The ratio between the total charge (summed over the three PMTs) and the total amplitude (similarly summed) is an effective pulse-shape discrimination (PSD) parameter, particularly for distinguishing between short Cherenkov pulses (low $Q/A$) and scintillation signals produced via UCN capture in the gas (large $Q/A$). A 2D histogram of the PMT-summed amplitude against $Q/A$ is shown in Fig.\@ \ref{fig:psa}. The red contour borders the bins containing events that survive all the cuts. A number of other features can be observed: the low-amplitude tail consists of activation events and ``edge'' UCN events where the proton and triton reach the entrance foil; pile-up events can be identified above and to the right of the red contour; and the sparse vertical band to the contour's immediate left is characteristic of Cherenkov pulses.

Cuts 1 and 5--7 are defined based on the charges of individual PMT waveforms, which for background tend to be relatively low. The relative-charge variables $D_1$, $D_2$, and $D_3$ provide crude localisation of events within the detector. For example, Cherenkov radiation released in a quartz window is typically concentrated on the associated PMT, giving a very large $q_i/Q$ (small $D_i$) and small $q_j/Q$ (large $D_j$) for $i \neq j$. UCNs principally form Gaussian-like distributions in these variables (centred on $\sim 0.7$), whilst activation events are strongly peaked towards higher values ($\sim 0.9$). Overall, the cuts select an average of $74.2$\,\% beam-on events and $1.88$\,\% beam-off background events. The number of background events in beam-on data before selection is estimated to be 9 times larger than the Poisson error of the UCN count; after selection it is over 4 times smaller.

Lastly, a degradation in the performance of the GADGET detector over the 44 data-taking days was observed, most prominently in the charge and amplitude distributions whose means drifted towards lower values at a rate of 0.7--0.8\,\% per day. This was likely caused by outgassing from the detector walls or small leaks that allowed air to enter the chamber, both of which can pollute the gas mixture and decrease the scintillation efficiency. To compensate for this effect and ensure that the selection cuts remain effective across the entire dataset, the beam-on distributions (of parameters used for selection) of each cycle are standardised and then re-scaled to the mean and standard deviation of the beam-on reference sample from which the cuts were derived. The remaining shape bias of the distributions between any two cycles after re-scaling has a negligible effect on the analysis.

\section{Disappearance Probability Computation}

The oscillation analysis relies on being able to determine the neutron disappearance probability for a given combination of magnetic field configuration (coil current and external field) and values of $\delta m$ and $\tau_{nn'}$. The disappearance probability depends on $\tau_{nn'}$ simply as $\tau_{nn'}^{-1}$, but its dependence on other parameters must be computed by numerically solving the Schr\"{o}dinger equation. We first build a model of the solenoid to compute the magnetic field at any point within the guide. It is modelled as the sum of 4092 (3995 for the main coil and 97 for one corrective coil) circular current loops. The radius $\rho$ and position $Z$ along the coil axis $z$ of each loop is known from the coil geometry. The field generated by each loop is given by the well established formula:

\begin{equation}
\vec B(\vec x)=\frac{\mu_0I}{4\pi\tilde\rho}\left(
\begin{array}{c}
\frac{z-Z}{\rho^2}\left[ \frac{2-\xi^2}{1-\xi^2}E(\xi)-2K(\xi) \right]x \\
\frac{z-Z}{\rho^2}\left[ \frac{2-\xi^2}{1-\xi^2}E(\xi)-2K(\xi) \right]y \\
\left[ \frac{(1+R/\rho)\xi^2-2}{1-\xi^2}E(\xi)+2K(\xi) \right] \\
\end{array}
\right)\:,
\end{equation}

where $E(\xi)$ and $K(\xi)$ are the complete elliptic integrals of the first and second kind, respectively, $\rho=\sqrt{x^2+y^2}$, $\tilde\rho=\sqrt{(z-Z)^2 +(R+\rho)^2}$, and $\xi^2=\frac{4R\rho}{\tilde\rho}$. The model describes the mapping along the solenoid axis at better than 0.2\% precision. The external magnetic field measured independently is then added to give the final field prediction. Fluctuations of the external field along the $z$-axis of a few of \si{\micro\tesla} were recorded during the experiment, which according to the simulation are analogous to shifts in current $\delta I(\textrm{A})=-2.5 \times 10^{-10} B(\textrm{T})$.

Ultracold neutron trajectories are simulated using the STARucn\@ \cite{starucn} Monte Carlo software, taking as an input the neutron velocity spectrum at PF2. It was verified that after $\mathcal{O}$(1)\,\si{metre} within the guide, the neutron velocities are independent of the initial spectrum, meaning related systematic uncertainties can be neglected. In the simulation, the neutron trajectories between two collision points within the guide are split into discrete segments. Knowing the position of the initial and final points of each segment $s$, as well as the neutron velocity, one can determine the total time $T(s)$ that the neutron spends in the segment and the instantaneous magnetic field $B(t)$ it experiences at each time \mbox{$t \in [0, T(s)]$}. The oscillation probability for each segment is obtained by solving the Schr\"{o}dinger equation using a standard Runge-Kutta algorithm (from the GSL library\@ \cite{gsl}):

\begin{equation}
i\frac{d}{dt}
\left(\begin{array}{c} \psi_{n} \\ \psi_{n'} \end{array}\right) 
=
\left(\begin{array}{cc} -\delta m+\mu_n B(t) & \tau_{nn'}^{-1} \\\tau_{nn'}^{-1} & \delta m-\mu_n B(t)  \end{array}\right) 
\left(\begin{array}{c} \psi_n \\ \psi_{n'} \end{array}\right)\:,
\end{equation}

where $\psi_n$ ($\psi_n'$) is the neutron (hidden neutron) wavefunction. As the probabilities over each segment are very small, the probability over a trajectory is the sum of the segment probabilities, and the overall probability is obtained by averaging over $n_t=50$ trajectories, corresponding to $453$ segments:

\begin{equation}
p(\delta m, I) = \frac{1}{n_t}\sum_{i=1}^{n_t} \sum_{s(i)} \left|\psi_{n'}(T(s))\right|^2 \:,
\end{equation}

where the index $i$ runs over the number of trajectories. Cross-checks in which the probability is averaged over an order-of-magnitude more trajectories indicate that fluctuations on 50-trajectory-averaged probabilities are on the order of 2\,\%. 

\begin{figure}
    \centering
    \includegraphics[width=0.99\linewidth]{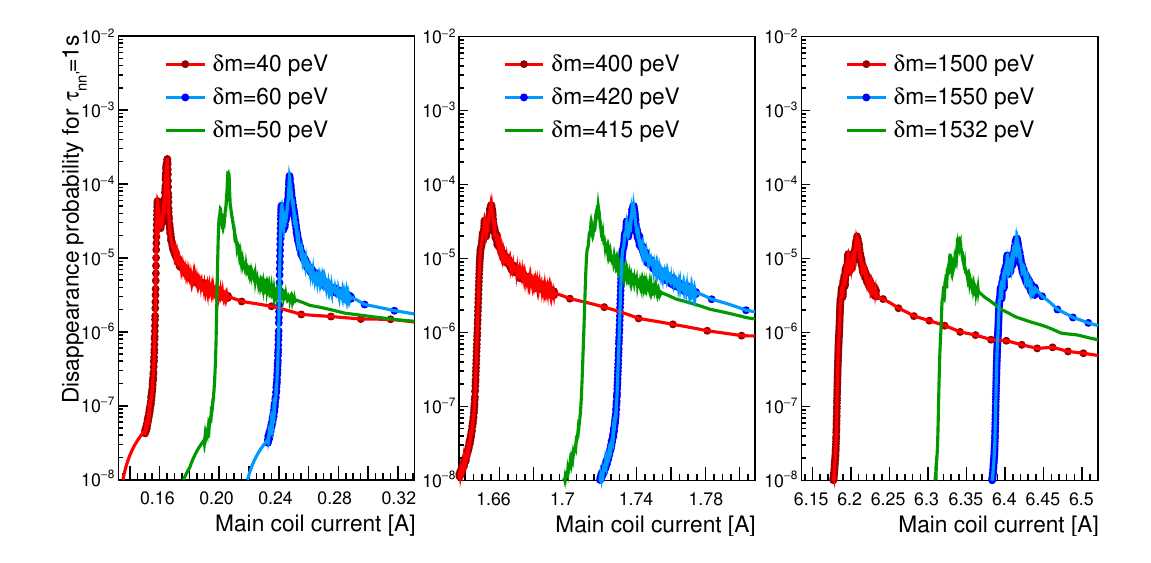}
    \vspace{-2mm}
    \caption{Numerical calculations of the disappearance probability as a function of coil current for two different $\delta m$ values (red and blue points), the interpolation at those values (red and blue curves), and the interpolation at an intermediate value (green curve), for three distinct $\delta m$ regimes.} 
    \label{fig:proba}
    \vspace{-4mm}
\end{figure}

As this computation is extremely time-consuming, it is not used directly to calculate the disappearance probabilities across the entire parameter space. Instead, it is performed for 55 values of $\delta m$ from 40 to 1900\,\si{\pico\electronvolt}, with $\tau_{nn'}=1$\,s. For each $\delta m$, the probability is computed for approximately 800 values of coil current in the range where the probability is most significant. A finer meshing of computed currents is employed in regions of maximum probability. These calculations are then used to interpolate the probability in current and $\delta m$. The computed and interpolated probabilities for three regions of $\delta m$ are plotted in Fig.\@ \ref{fig:proba}. For each region, two neighbouring sets of simulated data in $\delta m$ space are shown (red and blue dots), as well as the interpolation for these specific $\delta m$ (red and blue curves) and an interpolation for an intermediate $\delta m$ (green curve).

\section{Oscillation Analysis}

As mentioned in section \ref{sec:data-taking}, for most data-taking cycles (approximately 68\,\%) the UCN beam was shared with another experiment, switching between the two ports every 200\,\si{\second}. This led to a small non-linear build-up during counting period $A$, despite the allotted stabilisation time, as the UCN flux returned to its peak. An adjusted ratio is computed to account for this effect, defined as:

\begin{equation}
\tilde R_{ABC}=\frac{N_B}{\alpha N_A + N_C}\:,
\end{equation}

where the factor \mbox{$\alpha=1.00035(5)$} is determined by fitting the UCN rates (of cycles following beam switches) with the function \mbox{$f(t) = p_0 + p_1 e^{-p_2t}$} and taking $\alpha$ as the weighted mean of:

\begin{equation}
\alpha_c = (t_{A_{1}} - t_{A_{0}})\,p_0\left[\int^{t_{A_{1}}}_{t_{A_{0}}} f(t)\,dt\right]^{-1} \:,    
\end{equation}

where $t_{A_{0}}$ and $t_{A_{1}}$ are temporal limits of the $A$ period and $c$ refers to an individual cycle. For cycles that do not immediately follow a beam switch, $\alpha$~is set to unity.


The distribution of $\tilde R_{ABC}$ is compatible with a normal distribution of mean unity and standard deviation \mbox{$\sigma_{\tilde R} = 8.64(7)\times10^{-4}$}. However, compared to the average value of the $\tilde R_{ABC}$ errors, $\left<\Delta \tilde R_{ABC}\right>=3.416(1)\times 10^{-4}$, the larger-than-expected dispersion of points around the mean indicates the presence of non-statistical (i.e.\@ non-Poisson) fluctuations in the data. This was also observed in the previous experiment~\cite{saenz} and originates from fluctuations of either the reactor power or temperature of the cold source at short timescales. Consequently, the errors on $\tilde R_{ABC}$ are enlarged by a common factor $\frac{\sigma_R}{\left<\Delta \tilde R_{ABC}\right>}=2.53$, which constitutes the dominant systematic in this analysis.

The final dataset comprises values of $\tilde R_{ABC}$ and associated errors $\Delta \tilde R_{ABC}$, plus corresponding values of main-solenoid current and external magnetic field along the $z$ axis ($\delta B$), which is converted into a current fluctuation giving a total effective current $I_B$. For each cycle, the expected value of $\tilde R_{ABC}$ is connected to the disappearance probability ($p$) at currents $I_B$, $I_A=I_B-\delta I$ and $I_C=I_B+\delta I$ as:

\begin{equation}
R(I_B)=\frac{2-2p(I_B)}{2-p(I_A)-p(I_C)}\:.
\end{equation}

For a fixed $\delta m$, we define a chi-square function:

\begin{equation}
\chi^2(\tau_{nn'},\delta B^+, \delta B^-)=\sum_k\left(\frac{\tilde R_{ABC}^k-R(I_k)}{\Delta \tilde R_{ABC}^k}\right)^2+\left(\frac{\delta B^-}{\Delta^-}\right)^2+\left(\frac{\delta B^+}{\Delta^+}\right)^2 \:,
\end{equation} 

where the fluctuations relevant to the reference magnetic field in both current orientations, $\delta B^+$ and $\delta B^-$, are included as Gaussian nuisance parameters with a precision of $\Delta^+=\Delta^-=2$\,\si{\micro\tesla}. The chi-square is then profiled over the nuisance parameters:

\begin{equation}
\chi^2_{\text{pr}}(\tau_{nn'})=\min_{\delta B^+, \delta B^-}\chi^2\left(\tau_{nn'},\delta B^+,\delta B^-\right) \:.
\end{equation}

The profiled chi-square obtained for the no-oscillation null hypothesis is $\chi^2_{\text{pr}}/\textrm{ndf}=8076.7/7964$. Allowing oscillations, a best-fit value of \mbox{$\chi^2_{\text{pr}}/\textrm{ndf}=8062.9/7964$} is found at \mbox{$\delta m=838$\,\si{\pico\electronvolt}} and \mbox{$\tau_{nn'}=117$\,\si{\milli\second}}. This local significance of $3.7\sigma$ must be evaluated against the look-elsewhere effect \cite{lookelsewhere}. As the analysis covers a current range of around $6$\,\si{\ampere}, whereas an oscillation signature occurs within $60$\,\si{\milli\ampere} (i.e.\@ the difference in current between the $A$ and $C$ periods), a rough estimate of the Bonferroni factor is given by the ratio of the two, which reduces the significance by a factor of 100. Therefore, it is concluded that the data reveals no evidence of neutron-to-hidden neutron oscillations and a limit on the mixing parameters can be set.

\begin{figure}
    \centering
    \includegraphics[width=0.95\linewidth]{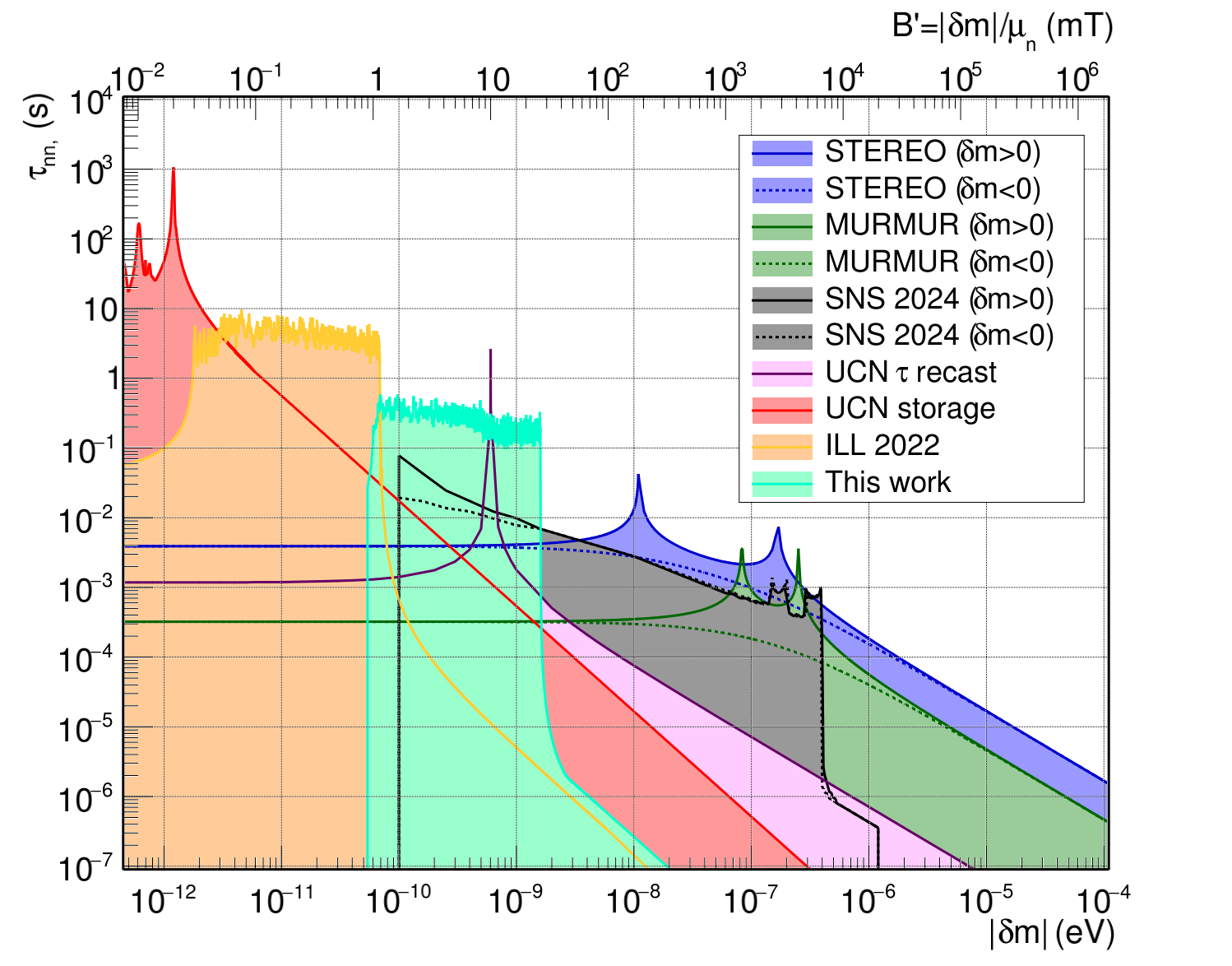}
    \vspace{-2mm}
    \caption{95\% CL limits in the $\delta m$--$\tau_{nn'}$ plane for this work (teal) and previous works:\@ \cite{Ban2007,Altarev2009,Berezhiani2012v2,Berezhiani2018,Ayres2022} (red),\@ \cite{Hostert2023} (purple),\@ \cite{Stasser2021} (green),\@ \cite{Almazn2022} (blue) ,\@ \cite{Gonzalez2024} (gray), and \@ \cite{saenz} (orange).}
    \label{fig:limit}
    \vspace{-4mm}

\end{figure}

The 95\% confidence level (CL) limit on $\tau_{nn'}$ is obtained by numerically solving $\chi^2_{\text{pr}}\left(\tau_{nn'}^{95\%}\right)-\min\chi^2(\tau_{nn'})=3.84$. In terms of systematic uncertainties, the effect of the magnetic field measurement uncertainty is of the order of a few percent on the limit. The numerical precision of 2\,\% on the probability calculation has been studied by shifting all Monte Carlo points by independent random Gaussian shifts of this size and recomputing the limit. As the observed variations are smaller than the magnetic effects by an order of magnitude, the analysis is fully constrained by the statistical and non-statistical fluctuation of $\tilde R_{ABC}$. The final exclusion contour in the $\delta m$--$\tau_{nn'}$ plane is given in Fig.\@ \ref{fig:limit}, alongside constraints obtained by previous experiments. As in Ref.\@ \cite{saenz} and despite the widening from the field gradient, the large fluctuations at the upper edge of the contour are caused by the smallness of the resonance width compared to the scanning step size.

\section{Conclusion}

The results presented in this work extend the search for neutron-to-hidden-neutron oscillations reported in Ref.\@ \cite{saenz}, by exploring the higher mass splitting region from 60 to 1600\,\si{\pico\electronvolt}, using a modified experimental setup. Using data collected at the ILL PF2 source from May to July 2024, oscillations in this higher mass regime are excluded with a sensitivity an order-of-magnitude lower than the previous experiment, due to the magnetic gradient necessary to widen the transition resonance. Conservative estimates of the allowed parameter-space region can be taken from the lowest common $\tau_{nn'}$ points of the limit and summarised as $\tau_{nn'} > 200$\,\si{\milli\second} for $|\delta m| \in [60, 400]$\,\si{\pico\electronvolt} and $\tau_{nn'} > 100$\,\si{\milli\second} for $|\delta m| \in [400, 1550]$\,\si{\pico\electronvolt}, which covers a significant part of the previously unconstrained parameter space. This type of direct disappearance measurement has likely reached its practical limits here, as the width of the resonance would require very long scans at higher fields. Moreover, covering higher mass splitting with higher fields would require more complex coils with either room-temperature magnets plus cooling systems or superconducting coils.

\section*{Acknowledgements}
We would like to acknowledge the technical support provided by T.\@ Brenner at ILL and D.\@ Etasse and D.\@ Goupillère at LPC Caen during all stages of the experiment. We would also like to thank G.\@ Tastevin and P.-J.\@ Nacher from LKB Paris for providing the solenoid at the heart of the experiment, as well as M.\@ Guigue at LPNHE Paris for fruitful discussions in the preparation of the experiment.

bibliographystyle{elsarticle-num}

\end{document}